\def\be{\begin{equation}}
\def\te{\end{equation}}
\def\ba{\begin{eqnarray}}
\def\nn{\nonumber}
\def\ta{\end{eqnarray}}

\newskip\humongous \humongous=0pt plus 1000pt minus 1000pt

\newif\ifdtup

\def\ha{{1\over 2}}

\def\(#1){(\ref{#1})}

\documentstyle[prd,aps]{revtex}

\newcommand{\kl}[3]{\mbox{$\rm #1$}^{\mu\nu , \alpha\beta}_{#2}(#3)}

\begin{document}
\draft
\title{Fluctuations in a Thermal Field and Dissipation of a Black Hole
Spacetime: Far-Field Limit}
\author{Antonio Campos}
\address{Center for Theoretical Physics, 
         Laboratory for Nuclear Science,
         Massachusetts Institute of Technology, 
         Cambridge, MA 02139, USA.}
\author{B. L. Hu}
\address{Department of Physics,
         University of Maryland,
         College Park, MD 20742, USA.}
\author{-- Talk by BLH in Bariloche, Argentina, Jan 1998. --\\
        Report umdpp \#98-128/MIT-CTP 2761}
\date{\today}
\maketitle
\begin{abstract}
{\scriptsize
We study the back reaction of a thermal field in a weak gravitational
background depicting the far-field limit of a black hole enclosed in
a box by the Close Time Path (CTP) effective action and the influence 
functional method. We derive the noise and dissipation kernels of 
this system in terms of quantities in quasi-equilibrium, and formally
prove the existence of a Fluctuation-Dissipation Relation (FDR) at all
temperatures between the quantum fluctuations
of the thermal radiance and the dissipation of 
the gravitational field. This dynamical self-consistent 
interplay between the quantum field and the classical spacetime is,
we believe, the correct way to treat back-reaction problems.
To emphasize this point we derive an Einstein-Langevin equation 
which describes the non-equilibrium dynamics
of the gravitational perturbations under the influence of the thermal
field. We show the connection between our method and the linear
response theory (LRT), and indicate how the functional method
can provide more accurate results than prior derivations of FDRs via LRT
in the test-field, static conditions. This method is in principle 
useful for treating fully non-equilibrium cases such as back reaction
in black hole collapse.}
\end{abstract}
\baselineskip=15pt


\section{Introduction}
\label{sec:intro}


In a recent essay \cite{Vishu} one of us outlined the program of black 
hole fluctuations and back reaction we are pursuing using stochastic
semiclassical gravity \cite{ssg} theory based on the Schwinger-Keldysh 
effective action \cite{ctp} and the Feynman-Vernon influence functional 
\cite{if} methods.
We mentioned prior works for static, quasi-static and dynamic black 
hole spacetimes and commented on how to improve on their shortcomings. 
In this paper, following the cues suggested in that essay, we discuss 
first the static case of a Schwarzschild black hole, with a further 
simplificaton of taking the far-field limit, and derive the
fluctuation-dissipation relation (FDR)\cite{FDR} under these 
conditions. In our view (following Sciama's \cite{Sciama} hints) 
the FDR embodies the back reaction of Hawking radiance \cite{Hawking}. 
In a recent work \cite{CamHu} we treated a relativistic thermal plasma
in a weak gravitational field. Since the far field limit 
of a Schwarzschild metric is just the perturbed Minkowski spacetime, 
the results there are useful for our present problem.
The more complicated case of near-horizon limit 
is also doable by calculating the fluctuations of the energy momentum 
tensor for the quantum field (with the help of e.g., the Page 
approximation \cite{Page}). 
Derivation of the dissipation and noise kernels for a static black hole 
in a cavity are currently under investigation \cite{CHR,HPR}.

\subsection{Fluctuation and Back reaction in Static Black Holes}

We recapitulate what was said before \cite{Vishu} on the state of art 
for problems in this case. Back reaction in this context 
usually refers to seeking a consistent solution of the 
semiclassical Einstein equation for the geometry of a black hole in 
equilibrium with its Hawking radiation 
(enclosed in a box to ensure relative stability).   
Much effort in the last 15 years has been devoted to finding 
a regularized energy-momentum for the back-reaction calculation. 
(See \cite{AndHis,JMO} for  recent status and earlier references.)  
Some important early work on back reaction was carried out by Bardeen, 
Hajicek and Israel \cite{BHI}, and York \cite{York}, and more recently 
by Massar, Parentani and Piran \cite{MPP} along similar lines.

Since the quantum field in such problems is assumed to be in a 
Hartle-Hawking state, concepts and techniques from thermal field 
theory are useful. Hartle and Hawking \cite{HarHaw}, 
Gibbons and Perry \cite{GibPer} used the periodicity condition of the
Green function on the Euclidean section to give a simple derivation 
of the Hawking temperature for a Schwarzschild black hole. The
most relevant work to our present problem is that by
Mottola \cite{Mot}, who showed that in some generalized 
Hartle-Hawking  states a FDR exists between the expectation values of 
the commutator and anti-commutator of the energy-momentum tensor.
This FDR has a form familiar in linear response theory \cite{LRT}:
\begin{equation}
   N_{abcd}(x, x')
        \ = \ \int_{-\infty}^{\infty}\frac{d\omega}{2\pi}
               e^{-i\omega(t-t')}\coth \left(\ha \beta \omega
                                       \right) 
               \tilde{D}_{abcd}({\bf x}, {\bf x'};\omega),
   \label{fdr}
\end{equation}
where $N$ and $D$ are the anticommutator and commutator functions of 
the energy-momentum tensor, respectively ($\tilde{D}$ is the temporal 
Fourier transform of $D$). That is,
\begin{eqnarray}
   N_{abcd}(x, x') 
        & = & \langle\{\hat{T}_{ab}(x),\,\hat{T}_{cd}(x')\}\rangle_{\beta} 
              \nn \\
   D_{abcd}(x, x') 
        & = & \langle[\hat{T}_{ab}(x),\,\hat{T}_{cd}(x')]\rangle_{\beta}.
\end{eqnarray}
He also identifies the two-point function $D$ as a dissipation
kernel by relating it to the time rate of change of the energy density
when the metric is slightly perturbed. Thus, Eq.(\ref{fdr}) represents
a bonafide FDR relating the fluctuations of a certain quantity 
(say, energy density) to the time rate of change of the very same quantity.

However, this type of FDR has rather restricted significance as it is
based on the assumption of a specific background spacetime (static in this
case) and state (thermal) of the matter field(s).
It is not adequate for the description of back reaction where the spacetime
and the state of matter are determined in a self-consistent manner by
their dynamics and mutual influence.  We should therefore look
for a FDR for a parametric family of metrics (belonging to a general class)
and a more general state of the quantum matter (in particular, for
Boulware and Unruh states). We expect the derivation of such a
FDR will be more complicated than the simple case above where the Green
functions are periodic in imaginary time throughout (not just as an initial
condition), and where one can simply take the results of linear response 
theory in thermal equilibrium (for all times) almost verbatim. 

Even in this simple case, it is worthwhile to note that there is a small
departure from standard linear response theory for quantum systems. This
arises from the observation that the dissipation kernel 
in usual linear response analyses is given by a two-point
commutator function of the underlying quantum field, which is independent
of the quantum state for free field theory. In this case, we are still
restricted to free fields in a curved background. However, since the
dissipation now depends on a two-point function of the stress-tensor,
it is a four-point function of the field, with appropriate derivatives and
coincidence limits. This function is, in general, state-dependent.
We have seen examples from related cosmological back-reaction problems
\cite{ssg} where it is possible to explicitly relate the dissipation
to particle creation in the field, which is definitely a
state-dependent process. For the
black-hole case, this would imply a quantum-state-dependent damping of
semiclassical perturbations.  To obtain a causal FDR for states more 
general than the Hartle-Hawking state, one needs to use the in-in 
(or Schwinger-Keldysh) formalism applied to a class of quasistatic metrics
(generalization of York \cite{York}) and calculate the fluctuations 
of the energy mometum tensor for the noise kernel. In our problem
such a calculation with back reaction is carried out in full detail, 
albeit only for a weak gravitational field here which depicts the far-field 
limit of a Schwarzschild black hole spacetimes.  We wish to address
the thermal field aspects of the problem, while saving the geometric
aspects in the near-horizon case for a later investigation.

\subsection{Thermal Fields in  Linear Gravity}

The behavior of a relativistic quantum field at finite temperature
in a weak gravitational field has been studied before by Gross, Perry 
and Yaffe \cite{GPY82}, Rebhan and coworkers \cite{Reb91}, 
de Almeida, Brandt, Frenkel and Taylor \cite{ABF94} 
for scalar and abelian gauge fields. In these work, the thermal 
graviton polarization tensor and the
effective action have been calculated and applied to the
study of the stability of hot flat/curved spaces and "dynamics"
of cosmological perturbations. To describe screening effects and 
stability of thermal quantum gravity, one needs only the real part of the
polarization tensor, but for damping effects, the imaginary part is
essential. The gravitational polarization tensor obtained from the
thermal graviton self-energy represents only a part (the thermal
correction to the vacuum polarization) of the finite temperature
quantum stress tensor. There is in general also contributions
from particle creation (from vacuum fluctuations at zero and finite 
temperatures). These processes engender dissipation in the dynamics of the 
gravitational field and their fluctuations appear as noise in 
the thermal field. We aim at finding a relation, the FDR, between these
two processes, which embodies the back reaction self-consistently.

In this work we use open system concepts and functional methods a la 
Schwinger-Keldysh \cite{ctp} and Feynman-Vernon \cite{if}. 
By casting the effective action in the form of an influence functional
we derive the noise and  dissipation  kernels explicitly and
prove that they satisfy a Fluctuation-Dissipation Relation (FDR) 
\cite{FDR} at all temperatures.
We also derive a stochastic semiclassical equation for the 
non-equilibrium dynamics of the gravitational field under the
influence of the thermal radiance.

We adopt the Hartle-Hawking picture where the black hole is bathed
eternally -- actually in quasi-thermal equilibrium --
in the Hawking radiance it emits. It will be described here 
by a massless scalar quantum field at the Hawking temperature.
As is well-known this quasi-equilibrium
condition is possible only if the black hole is enclosed in a
box of size slightly larger than the event horizon \cite{York} 
(or embedded in an anti-de Sitter space \cite{HawPag}).
In the asymptotic limit, the gravitational field is described by a linear
perturbation from Minkowski spacetime. In equilibrium  the thermal 
bath can be characterized by a relativistic fluid with a four-velocity
(time-like normalized vector field) $u^\mu$, 
and temperature in its own rest frame  $\beta^{-1}$. 
Taking into account the four-velocity $u^\mu$ of the 
fluid, a manifestly Lorentz-covariant approach to thermal field 
theory may be used \cite{Wel82}. However, in order to simplify the 
involved tensorial structure we work in the co-moving coordinate 
system of the fluid where $u^\mu = (1,0,0,0)$.

By making conformal transformations on the field and the spacetime,
our results may be easily generalized to the case of a conformally 
coupled quantum scalar field at finite temperature in a spatially 
flat Friedmann-Robertson-Walker universe \cite{Hu82}. Indeed we 
have earlier used the functional method and the Brownian motion 
paradigm \cite{if} to study similar problems in 
semiclassical gravity \cite{BD82}. We found that quantum noise
arising from fluctuations in the particle creation would constitute
a stochastic source (whose effect 
can overdominate the expectation value of the energy momentum tensor
in the semiclassical Einstein equation \cite{DTmn})
in a new form of Einstein-Langevin equation 
\cite{ssg}. We also came 
to the understanding that back reaction of vacuum quantum field 
processes (such as particle creation) on the dynamics of the early
universe near the Planck time is summarily a manifestation of a FDR 
in semiclassical gravity.

In Sec.~\ref{sec:effective action}, we describe our model and 
the derivation of the thermal CTP effective action. We compare it 
with the influence action \cite{if} and identify the dissipation and 
the noise kernels representing the linear response of the gravitational
field and the quantum fluctuations of the thermal radiance,
respectively. In Sec.~\ref{sec:FDR} we show that they obey a 
fluctuation-dissipation relation at all temperatures. In 
Sec.~\ref{sec:EL equation} we show how to derive a stochastic 
semiclassical equation for the gravitational perturbations from 
this effective action, which depicts the nonequilibrium dynamics 
of the gravitational field in a thermal radiation bath.


\section{CTP effective action at finite temperature}
\label{sec:effective action}


\subsection{The model}
 
In this section, we derive the CTP effective action for a thermal
quantum field in a classical gravitational background. 
To describe the radiation we consider a
free massless scalar field $\phi$ arbitrarily coupled to a
gravitational field $g_{\mu\nu}$ with classical action
\begin{equation}
   S_m[\phi,g_{\mu\nu}]
        \ = \ -{1\over2}\int d^nx\ \sqrt{-g}
               \left[ g^{\mu\nu}\partial_\mu\phi\partial_\nu\phi
                     +\xi(n) R\phi^2
               \right],
\end{equation}
where $R(x)$ is the scalar curvature and the arbitrary parameter $\xi (n)$
defines the type of coupling between the scalar field and the
gravitational field. If $\xi(n) = (n-2)/[4(n-1)]$, where $n$ is the
spacetime dimensions, the field is said to be conformally coupled; 
if $\xi(n) = 0$ the quantum field is said to be minimally coupled. In
the weak field limit we consider a small perturbation $h_{\mu\nu}$ from
flat spacetime $\eta_{\mu\nu}$
\begin{equation}
   g_{\mu\nu}(x) 
        \ = \ \eta_{\mu\nu} + h_{\mu\nu}(x), 
\end{equation}
with signature $(-,+,\cdots ,+)$ for the Minkowski metric. 
Using this metric and neglecting the surface terms that appear
in an integration by parts, the action for the scalar field may be written
perturbatively as 
\begin{equation}
   S_m[\phi,h_{\mu\nu}]
        \ = \  {1\over2}\int d^nx\ \phi
               \left[ \Box + V^{(1)} + V^{(2)} + \cdots
               \right] \phi,
\end{equation}
where the first and second order perturbative operators $V^{(1)}$ and 
$V^{(2)}$ are given by 
\begin{eqnarray}
   V^{(1)}
        & \ \equiv \ & - \left\{ [\partial_\mu\bar h^{\mu\nu}(x)]
                                 \partial_\nu
                                +\bar h^{\mu\nu}(x)\partial_\mu
                                 \partial_\nu
                                +\xi(n) R^{(1)}(x)
                          \right\},
                     \nonumber \\
   V^{(2)}
        & \ \equiv \ & \left\{ [\partial_\mu\hat h^{\mu\nu}(x)]
                               \partial_\nu
                              +\hat h^{\mu\nu}(x)\partial_\mu
                               \partial_\nu
                              -\xi(n) ( R^{(2)}(x)
                                    +{1\over2}h(x)R^{(1)}(x))
                       \right\}.   
\end{eqnarray}
In the above expressions, $R^{(k)}$ is the $k$-order term in the
pertubation $h_{\mu\nu}(x)$ of the scalar curvature and the
definitions $\bar h_{\mu\nu}$ and $\hat h_{\mu\nu}$ denote a
linear and a quadratic combination of the perturbation, respectively,
\begin{eqnarray}
   \bar h_{\mu\nu}
        & \ \equiv \ & h_{\mu\nu} - {1\over2} h \eta_{\mu\nu},
                     \nonumber \\
   \hat h_{\mu\nu}
        & \ \equiv \ & h^{\,\, \alpha}_\mu h_{\alpha\nu}
                      -{1\over2} h h_{\mu\nu}
                      +{1\over8} h^2 \eta_{\mu\nu}
                      -{1\over4} h_{\alpha\beta}h^{\alpha\beta} 
                       \eta_{\mu\nu}.
   \label{eq:def bar h}
\end{eqnarray}

For the gravitational field we take the following action
\begin{eqnarray}
   S^{div}_g[g_{\mu\nu}]
        & \ = \ & {1\over\ell^{n-2}_P}\int d^nx\ \sqrt{-g}R(x)
                \nonumber \\
        &       & +{\alpha\bar\mu^{n-4}\over4(n-4)}
                   \int d^nx\ \sqrt{-g}
                   \left[ 3R_{\mu\nu\alpha\beta}(x)
                           R^{\mu\nu\alpha\beta}(x)
                         -\left( 1-360(\xi(n) - {1\over6})^2
                          \right)R(x)R(x)
                   \right].
\end{eqnarray}
The first term is the classical Einstein-Hilbert action and the second
divergent term in four dimensions is the counterterm used in order to 
renormalize the effective action. In this action $\ell^2_P = 16\pi G$,
$\alpha = (2880\pi^2)^{-1}$ and $\bar\mu$ is an arbitrary mass scale.
It is noteworthy that the counterterms are independent of the
temperature because the thermal contribution to the effective action
is finite and does not include additional divergencies.

\subsection{CTP effective action}
\label{subsec:CTP eff act}

The CTP effective action at finite temperature for a free quantum scalar 
field in a gravitational background is given by 
\begin{equation}
   \Gamma^\beta_{CTP}[h^\pm_{\mu\nu}]
        \ = \ S^{div}_g[h^+_{\mu\nu}] 
             -S^{div}_g[h^-_{\mu\nu}]
             -{i\over2}Tr\{ \ln\bar G^\beta_{ab}[h^\pm_{\mu\nu}]\},
   \label{eq:eff act two fields}       
\end{equation}
where $ a, b = \pm$ denote the forward and backward time path and 
$\bar G^\beta_{ab}[h^\pm_{\mu\nu}]$ is the complete 
$2\times 2$ matrix propagator with 
thermal boundary conditions for the differential operator 
$\Box + V^{(1)} + V^{(2)} + \cdots$. Although the actual form of 
$\bar G^\beta_{ab}$ cannot be explicitly given, it is easy 
to obtain a perturbative expansion in terms of $V^{(k)}_{ab}$, the 
$k$-order matrix version of the complete differential operator 
defined by $V^{(k)}_{\pm\pm} \equiv \pm V^{(k)}_{\pm}$ and 
$V^{(k)}_{\pm\mp} \equiv 0$, and $G^\beta_{ab}$, the thermal matrix 
propagator for a massless scalar field in flat spacetime \cite{CamHu}.
To second order $\bar G^\beta_{ab}$ reads,
\begin{eqnarray}
   \bar G^\beta_{ab}
        \ = \  G^\beta_{ab}
              -G^\beta_{ac}V^{(1)}_{cd}G^\beta_{db}
              -G^\beta_{ac}V^{(2)}_{cd}G^\beta_{db}
              +G^\beta_{ac}V^{(1)}_{cd}G^\beta_{de}
               V^{(1)}_{ef}G^\beta_{fb}
              +\cdots
\end{eqnarray}
Expanding the logarithm and dropping one term independent of the
perturbation $h^\pm_{\mu\nu}(x)$, the CTP effective action may be
perturbatively written as,
\begin{eqnarray}
   \Gamma^\beta_{CTP}[h^\pm_{\mu\nu}]
        & \ = \ &  S^{div}_g[h^+_{\mu\nu}] - S^{div}_g[h^-_{\mu\nu}]
                \nonumber \\
        &       & +{i\over2}Tr[ V^{(1)}_{+}G^\beta_{++}
                               -V^{(1)}_{-}G^\beta_{--}
                               +V^{(2)}_{+}G^\beta_{++}
                               -V^{(2)}_{-}G^\beta_{--}
                              ]
                \nonumber \\
        &       & -{i\over4}Tr[  V^{(1)}_{+}G^\beta_{++}
                                 V^{(1)}_{+}G^\beta_{++}
                               + V^{(1)}_{-}G^\beta_{--}
                                 V^{(1)}_{-}G^\beta_{--}
                               -2V^{(1)}_{+}G^\beta_{+-}
                                 V^{(1)}_{-}G^\beta_{-+}
                              ]. 
   \label{eq:effective action}
\end{eqnarray}

In computing the traces, some terms containing divergencies are canceled 
using counterterms introduced in the classical gravitational action after
dimensional regularization. In general, the non-local pieces are of the
form $Tr[V^{(1)}_{a}G^\beta_{mn}V^{(1)}_{b}G^\beta_{rs}]$.
In terms of the Fourier transformed thermal propagators 
$\tilde G^\beta_{ab}(k)$ these traces can be written as,
\begin{equation}
   Tr[V^{(1)}_{a}G^\beta_{mn}V^{(1)}_{b}G^\beta_{rs}]
        \ = \  \int d^nxd^nx'\ 
               h^a_{\mu\nu}(x)h^b_{\alpha\beta}(x')
               \int {d^nk\over(2\pi)^n}{d^nq\over(2\pi)^n}
               e^{ik\cdot (x-x')}
               \tilde G^\beta_{mn}(k+q)\tilde G^\beta_{rs}(q)
               \kl{T}{}{q,k},
   \label{eq:trace}
\end{equation}
where the tensor $\kl{T}{}{q,k}$ is defined in \cite{CamHu}
after an  expansion in terms of a basis of 14 tensors \cite{Reb91}.
In particular, the last trace of (\ref{eq:effective action}) may be 
split in two different kernels $\kl{N}{}{x-x'}$ and $\kl{D}{}{x-x'}$,
\begin{equation}
   {i\over2}Tr[V^{(1)}_{+}G^\beta_{+-}V^{(1)}_{-}G^\beta_{-+}]
        \ = \ -\int d^4xd^4x'\ 
               h^+_{\mu\nu}(x)h^-_{\alpha\beta}(x')
               [   \kl{D}{}{x-x'}
                +i \kl{N}{}{x-x'}
               ].
\end{equation} 
One can express the Fourier 
transforms of these kernels, respectively, as
\begin{eqnarray}
   \kl{\tilde N}{}{k}
        & \ = \ & \pi^2\int {d^4q\over(2\pi)^4}\ 
                  \left\{ \theta(k^o+q^o)\theta(-q^o)
                         +\theta(-k^o-q^o)\theta(q^o)
                         +n_\beta(|q^o|)+n_\beta(|k^o+q^o|)
                  \right.
                \nonumber \\
        &       & \hskip2cm
                  \left. +2n_\beta(|q^o|)n_\beta(|k^o+q^o|)
                  \right\}\delta(q^2)\delta[(k+q)^2]\kl{T}{}{q,k},
   \label{eq:N}
\end{eqnarray}
\begin{eqnarray}
   \kl{\tilde D}{}{k}
        & \ = \ & -i\pi^2\int {d^4q\over(2\pi)^4}\
                  \left\{ \theta(k^o+q^o)\theta(-q^o)
                         -\theta(-k^o-q^o)\theta(q^o)
                         +sg(k^o+q^o) n_\beta(|q^o|)
                  \right.
                \nonumber \\
        &       & \hskip2cm
                  \left. -sg(q^o)n_\beta(|k^o+q^o|)
                  \right\}\delta(q^2)\delta[(k+q)^2]\kl{T}{}{q,k}.
   \label{eq:D}
\end{eqnarray}
Using the property $\kl{T}{}{q,k} = \kl{T}{}{-q,-k}$, it is easy to 
see that $\kl{N}{}{x-x'}$ is symmetric and $\kl{D}{}{x-x'}$ 
antisymmetric in their arguments; that is, $\kl{N}{}{x} = \kl{N}{}{-x}$ 
and $\kl{D}{}{x} = -\kl{D}{}{-x}$. 

To properly identify the physical meanings of these kernels
we have to write the renormalized CTP effective action at finite
temperature (\ref{eq:effective action}) in an influence functional form
\cite{if}. $\mbox{\rm N}$, the imaginary part of the CTP effective action 
can be identified with the noise kernel and $\mbox{\rm D}$, the 
antisymmetric piece of the real part, with the dissipation kernel.
In Sec.~\ref{sec:FDR} we will see that these kernels thus identified
indeed satisfy a thermal FDR. 

If we denote the difference and the sum of the perturbations
$h^\pm_{\mu\nu}$ defined along each branch $C_\pm$ of the complex time
path of integration $C$ by
$[h_{\mu\nu}] \equiv h^+_{\mu\nu} - h^-_{\mu\nu}$ and
$\{h_{\mu\nu}\} \equiv h^+_{\mu\nu} + h^-_{\mu\nu}$, respectively, 
the influence functional form of the thermal CTP effective action may 
be written to second order in $h_{\mu\nu}$ as,
\begin{eqnarray}
   \Gamma^\beta_{CTP}[h^\pm_{\mu\nu}]
        & \ \simeq \ & {1\over2\ell^2_P}\int d^4x\ d^4x'\
                       [h_{\mu\nu}](x)\kl{L}{(o)}{x-x'}
                       \{h_{\alpha\beta}\}(x')
                     \nonumber \\
        &            &+{1\over2}\int d^4x\ 
                       [h_{\mu\nu}](x)T^{\mu\nu}_{(\beta)}
                     \nonumber \\
        &            &+{1\over2}\int d^4x\ d^4x'\ 
                       [h_{\mu\nu}](x)\kl{H}{}{x-x'}
                       \{h_{\alpha\beta}\}(x')
                     \nonumber \\
        &            &-{1\over2}\int d^4x\ d^4x'\ 
                       [h_{\mu\nu}](x)\kl{D}{}{x-x'}
                       \{h_{\alpha\beta}\}(x')
                     \nonumber \\
        &            &+{i\over2}\int d^4x\ d^4x'\ 
                       [h_{\mu\nu}](x)\kl{N}{}{x-x'}
                       [h_{\alpha\beta}](x').
\end{eqnarray}
The first line is the Einstein-Hilbert
action to second order in the perturbation $h^\pm_{\mu\nu}(x)$. 
$\kl{L}{(o)}{x}$ is a symmetric kernel ({\sl i.e.} 
$\kl{L}{(o)}{x}$ = $\kl{L}{(o)}{-x}$) and its Fourier transform is
given by 
\begin{equation}
   \kl{\tilde L}{(o)}{k}
        \ = \ {1\over4}\left[ - k^2 \kl{T}{1}{q,k}
                              +2k^2 \kl{T}{4}{q,k}
                              + \kl{T}{8}{q,k}
                              -2\kl{T}{13}{q,k}
                       \right].
\end{equation}
The fourteen elements of the tensor basis $\kl{T}{i}{q,k}$ 
($i=1,\cdots,14$) are defined in \cite{Reb91}.
In the second line $T^{\mu\nu}_{(\beta)}$ has the form of a perfect fluid 
stress-energy tensor
\begin{equation}
   T^{\mu\nu}_{(\beta)}
        \ = \ {\pi^2\over30\beta^4}
              \left[ u^\mu u^\nu + {1\over3}(\eta^{\mu\nu}+u^\mu u^\nu)
              \right],
\end{equation}
where $u^\mu$ is the four-velocity of the plasma and the factor
${\pi^2\over30\beta^4}$ is the familiar thermal energy density for
massless scalar particles at temperature $\beta^{-1}$. In the third
line, the Fourier transform of the symmetric kernel $\kl{H}{}{x}$ can 
be expressed as
\begin{eqnarray}
   \kl{\tilde H}{}{k}
        & \ = \ &  -{\alpha k^4\over4}
                   \left\{ {1\over2}\ln {|k^2|\over\mu^2}\kl{Q}{}{k}
                          +{1\over3}\kl{\bar Q}{}{k}
                   \right\}
                \nonumber \\
        &       &  +{\pi^2\over180\beta^4}
                   \left\{ - \kl{T}{1}{u,k}
                           -2\kl{T}{2}{u,k}
                           + \kl{T}{4}{u,k}
                           +2\kl{T}{5}{u,k}
                   \right\}
                \nonumber \\
        &       &  +{\xi\over96\beta^2}
                   \left\{    k^2 \kl{T}{1}{u,k}
                           -2 k^2 \kl{T}{4}{u,k}
                           -      \kl{T}{8}{u,k}
                           +2     \kl{T}{13}{u,k}
                   \right\}
                \nonumber \\
        &       &  +\pi\int {d^4q\over(2\pi)^4}\
                   \left\{ \delta(q^2)n_\beta(|q^o|)
                           {\cal P}\left[ {1\over(k+q)^2}
                                   \right]
                          +\delta[(k+q)^2]n_\beta(|k^o+q^o|)
                           {\cal P}\left[ {1\over q^2}
                                   \right]
                   \right\}\kl{T}{}{q,k},
   \label{eq:grav pol tensor}
\end{eqnarray}
where $\mu$ is a simple redefinition of the renormalization parameter
$\bar\mu$ given by
$\mu \equiv \bar\mu \exp ({23\over15} +
{1\over2}\ln 4\pi - {1\over2}\gamma)$, and the tensors $\kl{Q}{}{k}$ 
and $\kl{\bar Q}{}{k}$ are defined, respectively, by
\begin{eqnarray}
   \kl{Q}{}{k}
        & \ = \ & {3\over2} \left\{               \kl{T}{1}{q,k}
                                    -{1\over k^2} \kl{T}{8}{q,k}
                                    +{2\over k^4} \kl{T}{12}{q,k}
                            \right\}
                \nonumber \\
        &       &-[1-360(\xi-{1\over6})^2]
                  \left\{               \kl{T}{4}{q,k}
                          +{1\over k^4} \kl{T}{12}{q,k}
                          -{1\over k^2} \kl{T}{13}{q,k}
                  \right\},
   \label{eq:Q tensor}
\end{eqnarray}
\begin{equation}
   \kl{\bar Q}{}{k}
        \ = \  [1+576(\xi-{1\over6})^2-60(\xi-{1\over6})(1-36\xi')]
                  \left\{               \kl{T}{4}{q,k}
                          +{1\over k^4} \kl{T}{12}{q,k}
                          -{1\over k^2} \kl{T}{13}{q,k}
                  \right\}. 
\end{equation}
In the above and subsequent equations, we denote the coupling
parameter in four dimensions $\xi(4)$ by $\xi$ and consequently 
$\xi'$ means $d\xi(n)/dn$ evaluated at $n=4$.
$\kl{\tilde H}{}{k}$ is the complete contribution of a free
massless quantum scalar field to the thermal graviton
polarization tensor\cite{Reb91,ABF94} and it is responsible for
the instabilities found in flat spacetime at finite temperature
\cite{GPY82,Reb91,ABF94}. \footnote{Note that the addition of
the contribution of other kinds of matter fields to the effective
action, even graviton contributions, does not change the tensor
structure of these kernels and only the overall factors are different
to leading order\cite{Reb91}}. Eq.~(\ref{eq:grav pol tensor}) 
reflects the fact that
the kernel $\kl{\tilde H}{}{k}$ has thermal as well as non-thermal
contributions. Note that it reduces to the first term in the zero
temperature limit ($\beta\rightarrow\infty$)
\begin{equation}
   \kl{\tilde H}{}{k}
        \ \simeq \ -{\alpha k^4\over4}
                     \left\{ {1\over2}\ln {|k^2|\over\mu^2}\kl{Q}{}{k}
                            +{1\over3}\kl{\bar Q}{}{k}
                     \right\}.
\end{equation}
and at high temperatures the leading term
($\beta^{-4}$) may be written as 
\begin{equation}
   \kl{\tilde H}{}{k}
        \ \simeq \ {\pi^2\over30\beta^4}
                    \sum^{14}_{i=1} 
                    \mbox{\rm H}_i(r) \kl{T}{i}{u,K},
\end{equation}
where we have introduced the dimensionless external momentum
$K^\mu \equiv k^\mu/|\vec{k}| \equiv (r,\hat k)$. The
$\mbox{\rm H}_i(r)$ coefficients were first given in \cite{Reb91} and
generalized to the next-to-leading order ($\beta^{-2}$) in
\cite{ABF94}. (They are given with the MTW sign convention  in 
\cite{CamHu})

Finally, as defined above, $\kl{N}{}{x}$ is the noise kernel
representing the random fluctuations of the thermal radiance and 
$\kl{D}{}{x}$ is the dissipation kernel, describing the 
dissipation of energy of the gravitational field.


\section{Fluctuation-Dissipation Relation and Linear Response Theory}
\label{sec:FDR}


\subsection{Fluctuation-Dissipation Relation}

These two kernels found above are functionally related by 
a fluctuation-dissipation relation (FDR). This relation
reflects the balance between the quantum
fluctuations in the thermal radiance and the energy loss by the
gravitational field. In \cite{CamHu} we have shown explicitly how 
this relation appears at zero temperature and at high temperature.
Here using the properties of the thermal propagators, 
we show that the FDR is formally satisfied for all temperatures.

We begin by showing that the 
FDR naively appears at the level of propagators as a direct 
consequence of the KMS relation \cite{KMS}. 
Then, using a generalization of this KMS relation, we see how the FDR 
is also satisfied by our noise and dissipation kernels.

To obtain the FDR at the level of propagators we need to introduce the
Schwinger and the Hadamard propagators. These propagators are defined
as the thermal average of the anticommutator
$G(x-x') 
 \equiv -i\langle [\phi(x),\phi(x')] \rangle_\beta$ and the commutator 
$G^{(1)}_\beta (x-x') 
 \equiv -i\langle \{\phi(x),\phi(x')\} \rangle_\beta$, respectively. 
The first represents the linear response of a relativistic system to
an external perturbation and the second the random fluctuations of the
system itself \cite{KMS,FDR}. Since we can write the KMS
condition satisfied by the propagators $G^\beta_{+-}$ and
$G^\beta_{-+}$ in Fourier space as
\begin{equation}
   \tilde G^\beta_{-+}(k)
        \ = \ e^{\beta k^o}\tilde G^\beta_{+-}(k),
   \label{eq:KMS}
\end{equation}
the Fourier transform of both the Schwinger $\tilde G(k)$ and the
Hadamard $\tilde G^{(1)}_\beta(k)$ propagators can be expressed, for
example, in terms of $\tilde G^\beta_{+-}(k)$ alone
\begin{eqnarray}
   \tilde G(k)
        & \ \equiv \ &  \tilde G^\beta_{-+}(k)
                       -\tilde G^\beta_{+-}(k)
          \    =   \    \left( e^{\beta k^o} - 1
                        \right) \tilde G^\beta_{+-}(k),
                     \nonumber\\
   \tilde G^{(1)}_\beta(k)
        & \ \equiv \ &  \tilde G^\beta_{-+}(k)
                       +\tilde G^\beta_{+-}(k)
          \    =   \    \left( e^{\beta k^o} + 1
                        \right) \tilde G^\beta_{+-}(k).
\end{eqnarray}
The FDR satisfied by these propagators follows inmediately from the
above equalities \cite{KMS,FDR}
\begin{equation}
   \tilde G^{(1)}_\beta(k)
        \ = \ \coth \left( {\beta k^o\over2}\right)\tilde G(k).
   \label{eq:FDR propagators} 
\end{equation}
Obviously, this relation can also be recovered if we write the
explicit expressions for the Fourier transform of the propagators
\begin{eqnarray}
   \tilde G(k)
        & \ = \ & -2\pi i\ sg(k^o) \delta(k^2),
                \nonumber\\
   \tilde G^{(1)}_\beta(k)
        & \ = \ & -2\pi i\ 
                   \coth \left({\beta|k^o|\over2 }\right)
                   \delta(k^2).
\end{eqnarray}
To use this last approach in our case could be a very difficult task
because one needs to compute the integrals for the noise and
dissipation kernels explicitly. On the other hand, if we follow the first
technique we only need to generalize the KMS condition of
Eq.~(\ref{eq:KMS}) to the product of two propagators.
This generalization reads
\begin{equation}
   \tilde G^\beta_{-+}(k+q)\tilde G^\beta_{+-}(q) 
        \ = \ e^{\beta k^o}
              \tilde G^\beta_{+-}(k+q)\tilde G^\beta_{-+}(q),
\end{equation}
and can be used to deduce the following formal identity
\begin{equation}
   \tilde G^\beta_{-+}(k+q)\tilde G^\beta_{+-}(q)
  +\tilde G^\beta_{+-}(k+q)\tilde G^\beta_{-+}(q)
        \ = \ \coth\left({\beta k^o\over2}\right)
              [ \tilde G^\beta_{-+}(k+q)\tilde G^\beta_{+-}(q)
               -\tilde G^\beta_{+-}(k+q)\tilde G^\beta_{-+}(q)].
   \label{eq:identity}
\end{equation}

Finally, one only needs to write, from the trace of Eq.~(\ref{eq:trace})
and the definitions (\ref{eq:N}) and (\ref{eq:D}), the noise and
dissipation kernels in terms of the propagators 
$\tilde G^\beta_{\pm\mp}$, respectively, as
\begin{equation}
   \kl{\tilde N}{}{k}
        \ = \  -{1\over4}\int {d^4q\over(2\pi)^4}\ 
                [ \tilde G^\beta_{-+}(k+q)\tilde G^\beta_{+-}(q)
                 +\tilde G^\beta_{+-}(k+q)\tilde G^\beta_{-+}(q)]
                \kl{T}{}{q,k},
   \label{eq:noise}
\end{equation}
\begin{equation}
    \kl{\tilde D}{}{k}
        \ = \  {i\over4}\int {d^4q\over(2\pi)^4}\ 
               [ \tilde G^\beta_{-+}(k+q)\tilde G^\beta_{+-}(q)
                -\tilde G^\beta_{+-}(k+q)\tilde G^\beta_{-+}(q)]
               \kl{T}{}{q,k},
   \label{eq:dissipation}
\end{equation}
and use the formal equality (\ref{eq:identity}) to prove
that they are related by the thermal identity
\begin{equation}
   \kl{\tilde N}{}{k} 
        \ = \ i\coth\left({\beta k^o\over2}\right)\kl{\tilde D}{}{k}.
\end{equation}
In coordinate space we have the analogous expression
\begin{equation}
   \kl{N}{}{x} 
        \ = \ \int d^4x'\ \mbox{\rm K}_{FD}(x-x')\kl{D}{}{x'},
\end{equation}
where the fluctuation-dissipation kernel $\mbox{\rm K}_{FD}(x-x')$ is
given by the integral
\begin{equation}
   \mbox{\rm K}_{FD}(x-x')
        \ = \ i \int {d^4k\over(2\pi)^4}\
                     e^{ik\cdot(x-x')}
                     \coth\left({\beta k^o\over2}\right).
\end{equation}
The proof of this FDR at finite temperature is in some sense formal
because we have assumed along the argument that the integrals are 
always well defined. Nevertheless, the exact results obtained for the
zero and high temperature limits \cite{CamHu} indicate that the noise and
dissipation kernels are well defined distributions \cite{Jon82}. The
asymptotic analysis has also been useful to determine the physical
origin of the fluctuations.

\subsection{Linear Response Theory}
\label{sec:LRT}

A fluctuation-dissipation relation is usually derived using
linear response theory (LRT). We now show the connection between the
LRT \cite{LRT} and the functional methods we have used here. In the 
spirit of LRT the gravitational field is considered as a weak 
external source which imparts disturbances to the radiance whose 
response is studied to linear order.

Let us first recall the main features of LRT. Consider a system 
described by the Hamiltonian operator $\hat H_o$ initially coupled 
linearly to an external driving agent, say $A_\alpha$. Since we are 
only interested in how the system responds to the external agent, 
and not the details of the agent, we will ignore the Hamiltonian 
for the external perturbation but write the complete operator 
Hamiltonian of the sytem as
\begin{equation}
   \hat H
        \ = \ \hat H_o + A_\alpha \hat J^\alpha, 
\end{equation}
where $J^\alpha$ is the current operator associated with the external 
agent. If the system is in thermal equilibrium before the external 
source is applied the first order response of the system to this 
external force is given by the thermal expectation value of the
commutator of the current operator over its thermal average
\begin{equation}
   \langle [J^\mu(x)-\langle J^\mu(x) \rangle_\beta,
            J^\nu(x')-\langle J^\nu(x') \rangle_\beta] 
 \rangle_\beta.
\end{equation}
In contrast, the intrinsic quantum fluctuations of the system are
described by the thermal average of the anticommutator. In our case, 
the conserved current operator is given by the stress-energy tensor 
$T^{\mu\nu}(x)$ as derived from the classical action. Our objective
is to show that the response and fluctuation
functions for the stress-energy tensor considered in the LRT are 
equivalent to the dissipation and noise kernels, respectively.

First, we write the classical action for the matter field to
linear order in the gravitational perturbations,
\begin{equation}
   S_m[\phi,h_{\mu\nu}]
        \ \simeq \ {1\over2}\int d^4x\ 
                   [ \phi\Box\phi + h_{\mu\nu}T^{\mu\nu} ],
\end{equation}
with the stress-energy tensor given by
\begin{equation}
   T^{\mu\nu}
        \ = \ P^{\mu\nu,\alpha\beta} 
              \partial_\alpha\phi\partial_\beta\phi
             +\xi \left( \eta^{\mu\nu}\Box - \partial^\mu\partial^\nu
                  \right) \phi^2.
\end{equation}
Note that $T^{\mu\nu}(x)$ is conserved if the classical unperturbed
equation of motion for $\phi$ is satisfied and it reduces to the 
stress-energy tensor for a scalar field in flat spacetime if 
$\xi = 0$. Alternatively, we can write the Hamiltonian formulation of
our problem. If we introduce the conjugate momentum variable of the
matter field to first order in the perturbation
\begin{equation}
   \Pi  
        \ \equiv \ {\partial {\cal L}\over\partial\dot\phi}
        \ \sim   \ \dot\phi 
                  +{1\over2}h_{\mu\nu}
                   {\partial T^{\mu\nu}\over\partial\dot\phi},
\end{equation}
the Hamiltonian can be written as
\begin{equation}
   H
        \ \simeq \  {1\over2}\int d^3\vec{x}
                    \left[ \Pi^2
                          +(\vec{\nabla}\phi)^2
                          - h_{\mu\nu}T^{\mu\nu}
                    \right]. 
\end{equation}
Note that to first order $\dot\phi$ and $\Pi$ are interchangeable in
the expression for the stress-energy tensor.

Using the thermal version of the Wick theorem \cite{Mil69,LeB96}, one 
can write, after some algebra, the equilibrium thermal average of the
two-point function for the stress-energy tensor operator at different 
spacetime points in terms of products of thermal propagators,
\begin{equation}
   \langle T^{\mu\nu}(x) T^{\alpha\beta}(x') \rangle_\beta
  -\langle T^{\mu\nu}(x) \rangle_\beta
   \langle T^{\alpha\beta}(x') \rangle_\beta
        =   -2\int {d^4k\over(2\pi)^4} e^{ik\cdot (x-x')}
                   \int {d^4q\over(2\pi)^4}\ 
                   \tilde G^\beta_{-+}(k+q)
                   \tilde G^\beta_{+-}(q)
                   \kl{T}{}{q,k}.
\end{equation}
Finally, defining
$ \Delta_\beta T^{\mu\nu}(x) \equiv  T^{\mu\nu}(x) 
 -\langle T^{\mu\nu}(x) \rangle_\beta$
and using the expressions for the noise and dissipation kernels
given in (\ref{eq:noise}) and (\ref{eq:dissipation}) respectively,
we obtain
\begin{equation}
   \langle \{ \Delta_\beta T^{\mu\nu}(x),
              \Delta_\beta T^{\alpha\beta}(x')
            \} 
   \rangle_\beta 
        \ = \  8\ \kl{N}{}{x-x'},
\end{equation}
\begin{equation}
   \langle [ \Delta_\beta T^{\mu\nu}(x),
              \Delta_\beta T^{\alpha\beta}(x')
            ] 
   \rangle_\beta 
        \ = \  8i\ \kl{D}{}{x-x'}.
\end{equation}
{From} these formal identities we conclude that the functional method
gives a description of the lowest order dynamics of a near-equilibrium
system equivalent to that given traditionally by the LRT.


\section{Einstein-Langevin equation}
\label{sec:EL equation}


To reinforce the points made at the beginning about the dynamic
nature of the back reaction of thermal radiance on the black hole
spacetime even for the quasi-static case, we now derive
from the thermal CTP effective action a dynamical equation 
governing the dissipative evolution
of the gravitational field under the influence of the
fluctuations of the thermal radiance. It is in the form of an
Einstein-Langevin equation \cite{ssg}.

We first introduce the influence functional \cite{if} 
${\cal F} \equiv \exp (iS_{IF})$ where the influence
action $S_{IF}$ is related to the CTP effective
action in the semiclassical limit by \cite{ssg},
\begin{equation}
   {\cal F}
        \ = \ \exp i\left( Re \{ \Gamma^\beta_{CTP}[h^\pm_{\mu\nu}] \}
                          +{i\over2}\int d^4x\ d^4x'\ 
                               [h_{\mu\nu}](x)\kl{N}{}{x-x'}
                               [h_{\alpha\beta}](x')
                    \right),
\end{equation}
where $Re\{\ \}$ denotes taking the real part. Following
\cite{if} we can interpret the real part of
the influence functional as the characteristic functional of a
non-dynamical stochastic variable $j^{\mu\nu}(x)$,
\begin{equation}
   \Phi([h_{\mu\nu}])
        \ = \ \exp \left( -{1\over2}\int d^4x\ d^4x'\ 
                           [h_{\mu\nu}](x)\kl{N}{}{x-x'}
                           [h_{\alpha\beta}](x')
                   \right).
   \label{eq:cf}
\end{equation} 
This classical stochastic field represents probabilistically the quantum
fluctuations of the matter field and is responsible for the
dissipation of the gravitational field. By definition, the
above characteristic functional is the functional Fourier transform of
the probability distribution functional ${\cal P}[j^{\mu\nu}]$ with
respect to $j^{\mu\nu}$,
\begin{equation}
   \Phi([h_{\mu\nu}])
        \ = \ \int {\cal D}j^{\mu\nu}\ {\cal P}[j^{\mu\nu}]\
              e^{i\int d^4x\ [h_{\mu\nu}](x)j^{\mu\nu}(x) }.   
   \label{eq:cf_pdf}
\end{equation}
Using (\ref{eq:cf}) one can easily see that the probability
distribution functional is related to the noise kernel by the formal
expression,
\begin{equation}
   {\cal P}[j^{\mu\nu}]
        \ = \ {\exp \left( -{1\over2}\int d^4x\ d^4x'\ 
                           j_{\mu\nu}(x)[\kl{N}{}{x-x'}]^{-1}
                           j_{\alpha\beta}(x')
                   \right)
               \over
               \int {\cal D}j^{\mu\nu}\
               \exp \left( -{1\over2}\int d^4x\ d^4x'\ 
                           j_{\mu\nu}(x)[\kl{N}{}{x-x'}]^{-1}
                           j_{\alpha\beta}(x')
                   \right)
              }.
   \label{eq:pdf}
\end{equation}
For an arbitrary functional of the stochastic field ${\cal E}[j^{\mu\nu}]$,  
the average value with respect to the previous probability distribution
functional is defined as the functional integral 
$\langle {\cal E}[j^{\mu\nu}] \rangle_j 
 \equiv \int {\cal D}[j^{\mu\nu}]\ 
             {\cal P}[j^{\mu\nu}] 
             {\cal E}[j^{\mu\nu}]$.
In terms of this stochastic average the influence functional can be
written as 
${\cal F} = 
 \langle \exp\left(i\Gamma^{st}_{CTP}[h^\pm_{\mu\nu}]
             \right) 
 \rangle_j$, where $\Gamma^{st}_{CTP}[h^\pm_{\mu\nu}]$ is the modified
effective action
\begin{equation}
   \Gamma^{st}_{CTP}[h^\pm_{\mu\nu}]
        \ \equiv \  Re \{ \Gamma^\beta_{CTP}[h^\pm_{\mu\nu}] \}
                   +\int d^4x\ [h_{\mu\nu}](x) j^{\mu\nu}(x).
   \label{eq:mea}
\end{equation}
Clearly, because of the quadratic nature of the characteristic 
functional (\ref{eq:cf}) and its relation with the probability 
distribution functional (\ref{eq:cf_pdf}), the field $j^{\mu\nu}(x)$ is a
zero mean Gaussian stochastic variable. This means that its two-point
correlation function, which is given in terms of the noise kernel by 
\begin{equation}
   \langle j^{\mu\nu}(x) j^{\alpha\beta}(x') \rangle_j
        \ = \ \kl{N}{}{x-x'},
   \label{eq:correlation}
\end{equation}
completely characterizes the stochastic process. The Einstein-Langevin
equation follows from taking the functional derivative of the
stochastic effective action (\ref{eq:mea}) with respect to 
$[h_{\mu\nu}](x)$ and  imposing $[h_{\mu\nu}](x) = 0$.
 In our case, this leads to
\begin{equation}
   {1\over\ell^2_P} 
   \int d^4x'\ \kl{L}{(o)}{x-x'} h_{\alpha\beta}(x') 
  +{1\over2}\ T^{\mu\nu}_{(\beta)}
  +\int d^4x'\ \left( \kl{H}{}{x-x'}
                     -\kl{D}{}{x-x'}
               \right) h_{\alpha\beta}(x')
  +j^{\mu\nu}(x)
        \ = \ 0.      
\end{equation}
To obtain a simpler and clearer expression we can rewrite this
stochastic equation for the gravitational perturbation in the harmonic
gauge $\bar h^{\mu\nu}_{\,\,\,\,\, ,\nu} = 0$,
\begin{equation}
   \Box\bar h^{\mu\nu}(x)
         + \ell^2_P
               \left\{ T^{\mu\nu}_{(\beta)}
                      +2P_{\rho\sigma,\alpha\beta}
                       \int d^4x'\ \left( \kl{H}{}{x-x'}
                                         -\kl{D}{}{x-x'}
                                   \right)\bar h^{\rho\sigma}(x')
                      +2j^{\mu\nu}(x)
               \right\} = 0,
\end{equation}
where we have used the definition for $\bar h^{\mu\nu}(x)$ written in 
(\ref{eq:def bar h}) and the tensor $P_{\rho\sigma,\alpha\beta}$ is
given by
\begin{equation}
   P_{\rho\sigma,\alpha\beta}
        \ = \ {1\over2}\left( \eta_{\rho\alpha}\eta_{\sigma\beta}
                             +\eta_{\rho\beta}\eta_{\sigma\alpha}
                             -\eta_{\rho\sigma}\eta_{\alpha\beta}
                       \right).
\end{equation} 
Note that this differential stochastic equation includes a non-local
term responsible for the dissipation of the gravitational field and a
noise source term which accounts for the fluctuations in the thermal 
radiance. They are connected by a FDR as described in the last section.
Note also that this equation in combination with the correlation for 
the stochastic variable (\ref{eq:correlation}) determine the two-point
correlation for the stochastic metric fluctuations 
$\langle \bar h_{\mu\nu}(x) \bar h_{\alpha\beta}(x') \rangle_j$
self-consistently.


\section{Conclusions}
\label{sec:conclusions}


In this paper we show how the functional methods can be used effectively
to study the non-equilibrium dynamics of a weak classical gravitational 
field in a thermal quantum field. The Close Time Path (CTP)
effective action and the influcence functional were used
to derive the noise and dissipation kernels of this system.
The back reaction of the thermal radiance on the gravitational field 
is embodied in a Fluctuation-Dissipation Relation (FDR), which connects
the fluctuations in the thermal radiation and the energy
dissipation of the gravitational field.
We prove formally the existence of such a relation for thermal fields
at all temperatures.

We also show the formal equivalence of this method with Linear Response Theory 
(LRT) for lowest order perturbance of a
near-equilibrium system, and how the response functions
such as the contribution of the quantum scalar field to the thermal 
graviton polarization tensor can be derived.
An important quantity not usually obtained in LRT but
of equal importance manifest in the CTP approach is the
noise term arising from the quantum and statistical fluctuations in the 
thermal field.

Finally, we emphasize that the back reaction is intrinsically
a dynamic process which traditional LRT calculations cannot capture
fully. We illustrate this point by deriving a Einstein-Langevin
equation for the non-equilibrium dynamics of the gravitational field with
back reaction from the thermal field. This method can be applied to 
quasi-dynamic \cite{Sciama,HRS}
or fully dynamic problems such as black hole collapse \cite{BHI,MPP,HLPS}.
To complete the present problem of quasi-static black hole 
back reaction, we need to perform the same calculation for the
full Schwarzschild spacetime. Currently we are working on the 
fluctuations of the energy-momentum tensor near the black hole 
horizon and the derivation of the noise kernel. Results will be 
reported in future publications.


\acknowledgments


This work was carried out when AC visited the University of Maryland
in Spring 1997 and finished in Spring 1998 at MIT while supported in 
part by the Comissionat per a Universitats i Recerca under a
cooperative agreement between MIT and Generalitat de Catalunya and 
funds provided by the U.S. Department of Energy (D.O.E.) under 
cooperative research agreement DE-FC02-94ER40818. BLH was supported 
in part by NSF grant PHY94-21849.
He thanks the organizers of the Workshop on
Quantum Gravity in the Southern Cone for their warm hospitality.
We have benifitted from discussions with Esteban Calzetta,
R. Martin, Alpan Raval, Sukanya Sinha and Enric Verdaguer.


\end{document}